# Pressure waves from air gun bubbles: a numerical analysis based on Finite Volume Method

Shi-Ping Wang (王诗平) [a, b], Hang Geng (耿航) [a, b], Shuai Zhang (张帅) [a, b*], Si-Wei Wang (王思伟) [c]

[a] *College of Shipbuilding Engineering, Harbin Engineering University, Harbin, 150001, China*
[b] *Nanhai Institute of Harbin Engineering University, Sanya, 572024, China*
[c] *Yichang Testing Technique Research Institute, Yichang, 443003, China*

**ABSTRACT:** The pressure wave emitted from the air gun contains many frequencies, among which the low-frequency waves are desirable for exploration and imaging, while the high-frequency waves need to be suppressed as they are harmful to marine species. The high-frequency waves originate from the fast oscillations of the flow during the release of the air, such as the impingement of the gas jet into the liquid, the expansion of the air gun bubble, and the interaction between the air gun body and the bubble. However, those dynamic and the emitted waves are adjustable by the special design of the air guns. To analyze the underlying relations, we present a numerical study with a compressible air gun bubble model using the volume of fluid (VOF) approach combined with the finite volume method (FVM) implemented in STAR-CCM+. The venting process of an air gun is investigated to reveal the influence of the air gun body. The results show that air gun pressure for the far field is mainly proportional to the expansion acceleration of the whole gas. Our results also indicate that the opening and chamber shape of the air gun affects the gas expansion acceleration, which influences the first peak of the pressure wave significantly. The larger the opening is, the faster the gas is released, the greater the amplitude of the first peak is. The larger the chamber length/diameter ratio, the slower the gas is released and the lower the amplitude of the first peak.

## 1. Introduction

The seismic air guns are the main acoustic wave source in ocean geological prospecting[1,2]. The air gun source is usually filled with high-pressure gas, which is instantly released into the ocean and creates an oscillating bubble emitting sound waves[3,4]. Such signals are over quite a wide range of frequencies and are reflected by various geological layers when propagating to the seafloor[5,6]. The principle can be seen in Fig. 1. However, the low frequencies (generally less than approximately 200 Hz) are helpful as they can penetrate the sea and the geologic layers of the seafloor and reflect from the target of interest[7]. The high frequencies are attenuated faster, cannot penetrate as deeply, and are of limited benefit to imaging[8]. What is more, certain studies suggest that the high frequencies in the air gun signals harm some marine life[9–11], apart from whales using low-frequency (15-100 Hz) sound for communication, is also considered the most susceptible to the nongeochemical functional high-frequency components of the air gun signals[12]. Therefore, there is a trend to improve existing air gun designs or create new ones to decrease the amplitude of the high-frequency component of the signals.

---

[*] Corresponding author.
   E-mail addresses: zhangshuai8@hrbeu.edu.cn

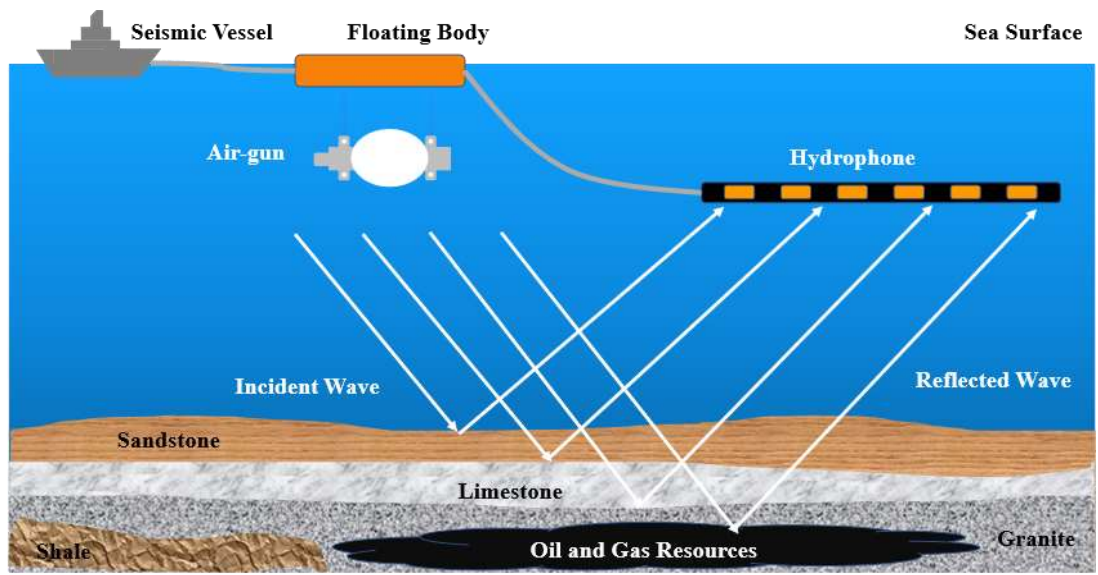

**Fig. 1**. Principle schematic diagram of the seismic air gun exploration.

Plenty of meaningful work has been done to reveal the spectral characteristics of the air gun source[13–16]. The Gilmore equation was first creatively applied to calculating the air gun's far-field pressure wavelet and altered the trial-and-error methods[17]. In which spherical, compressible bubble pulsation equations describe the bubble motion as it has high calculation efficiency[18,19]. Since then, many authors have used such a framework while more factors are considered, and some parameters are discussed according to experiments, such as the damping mechanisms resulting in attenuation of bubble oscillation[20–25] and the transfer of air between the chamber and the bubble[26,27]. However, all the above were based on the spherical bubble theory without regard to the body of the air gun. The presence of a structure[28–36] or free surface[37–43] can significantly cause non-spherical expansion and collapse of bubbles.

The generation and oscillation of the bubble are undoubtedly related to the air gun body accordingly, and lots of work has been done. Watson et al. proposed a quasi-1-D model of an air gun united with a spherical bubble, which considers gas dynamics and spatial pressure variation in the launch chamber to investigate the control of initial peaks of the source[7]. The boundary element method (BEM) is creatively applied to study the interaction between the bubble and the air gun body as the air gun body is simplified as a cylinder [44] or two cylinders[45], which is often used to study the bubble dynamics[45–53] as it takes into account non-spherical deformation even splitting[54] of the bubble surface. Nevertheless, it is much more difficult for the BEM to solve the venting process of the air gun, while the finite volume method (FVM) has a significant advantage. Based on VOF, Zhang et al. established a compressible two-phase flow model in Open FOAM[55,56], considering the effects of the air gun body, including the opening form and area.

The generation of the air gun bubble is essential but difficult to determine, which decides the initial peak and high-frequency pressure waves[1,7]. In previous studies, the initial air release process is always ignored[57,58] or adopted as an analytical solution[7,23,59]. It is hard to estimate the first pressure peak as the outflow of gas is uncontrolled and rapid after the opening of the air gun is opened. Hence, there needs to be more studies about how the air gun structure affects gas transfer and the initial peak, which are usually thought of as the source feature of the air gun.

This work proposes a compressible air gun bubble model, which is implemented in STAR-CCM+, and the effects of the air gun body could be considered. The venting process of the air gun is investigated using the FVM united with VOF, one of the standard interface-capturing methods. We analyze how the air gun chamber and opening affect gas transfer and the initial peak of the fluid pressure, which control the source signature and are associated with high-frequency acoustic waves. The study may provide a reference for air gun designers to inform future design and operation.

## 2. Methodology

*2.1. Numerical model*

The air gun bubble oscillation in the water is considered a two-phase gas-liquid flow[56]. This paper uses the FVM combined with VOF in STAR-CCM+ to solve this problem. The Segregated Fluid Isothermal model is used here as the short period of the bubble pulsation results in minimal temperature impact[27,56,60], and a constant temperature is set at 300 K for the whole field. The High-Resolution Interface Capturing (HRIC) scheme is applied to track the sharp interfaces by which the air and water always remain separated. The whole calculation domain is divided into cells where the sum of the volume fractions of the water and gas equals 1[5]. That means,

$$\alpha_w + \alpha_g = 1 \tag{1}$$

Where $\alpha_w$ means the volume fraction of water, $\alpha_g$ means the volume fraction of gas.

The material properties in the cell where the interface is located depend on the gas and water mixture properties.

$$\rho = \alpha_w \rho_w + \alpha_g \rho_g \tag{2}$$

Ignore the diffusion term, and the transport equation of water volume fraction is described as follows:

$$\frac{\partial}{\partial t}\int_V \alpha_w dV + \oint_A \alpha_w \mathbf{v} d\mathbf{a} = \int_V -\frac{\alpha_w}{\rho_w}\frac{D\rho_w}{Dt}dV \tag{3}$$

Where $A$ and $V$ are the area and volume of the cell, $\mathbf{v}$ indicates the mixture velocity (mass-averaged), $\mathbf{a}$ indicates the surface area vector, and $D\rho_w/Dt$ is the Lagrangian derivative of the water densities.

As the water volume fraction transport is solved, adjust the volume fraction of the air to ensure the sum of the volume fractions of the water and gas equals 1. The air gun bubble flow is inertia-controlled[27], and viscous influence is ignored as the associated Reynolds number is estimated as $O(10^6)$, so the total mass, momentum, and energy conservation equations for all phases are given:

$$\oint_A \rho \mathbf{v} d\mathbf{a} + \frac{\partial}{\partial t}\int_V \rho dV = 0 \tag{4}$$

$$\oint_A \rho \mathbf{v} \otimes \mathbf{v} \cdot d\mathbf{a} + \frac{\partial}{\partial t}(\int_V \rho \mathbf{v} dV) = -\oint_A p d\mathbf{a} \tag{5}$$

$$\oint_A (\rho H \mathbf{v} + p) \cdot d\mathbf{a} + \frac{\partial}{\partial t}\int_V \rho E dV = 0 \tag{6}$$

Where $E$ and $H$ are the total energy and enthalpy, respectively.

State Equations for air and water are also needed to solve the above equations. For gas, the ideal gas law is used here, a function of density, temperature, and pressure:

$$\rho_g = \frac{p_g}{RT}, R_g = \frac{R_u}{M} \tag{7}$$

Where $p_g$ is the gas pressure, $R_g$ is the gas constant, $R_u$ is the universal gas constant value 8314.4621, and $M$ is the molecular weight of the gas.

For water, the models of IAPWS-IF97 (International Association for the Properties of Water and Steam; Industrial Formulation, 1997) are adopted in this paper. The IAPWS-IF97 provides fundamental polynomial equations for the specific Gibbs free energy, which is compressible.

For boundary, it is better to choose the non-reflecting option to prevent spurious numerical reflection of the solution into the solution domain[61]. However, the model has no such option, and the overset mesh approach offers a good solution, as shown in Fig. 2 (a). The overset simulation has a background region enclosing the entire solution domain and the overset region of bubble size. The background region is large enough that the pressure wave has yet to reach the boundary before the end time, divided into sparse. In contrast, the grid spacing in the overset area is relatively large, mainly to improve computational efficiency. The boundary for the background region can be the pressure outlet. The accuracy and precision of the overset mesh are discussed next.

Since the air gun bubble oscillation in the water is considered a segregated flow, using the implicit unsteady solver is an excellent choice compared with the explicit unsteady one. For transient simulations in the equations above, the governing equations' solution is generally obtained at different time steps, whereby the solution at time $t$ requires the solutions of the previous time steps and the implicit time integration is adopted here. The second-order upwind (SOU) scheme is adopted to compute the convective flux. Furthermore, the gradients are calculated using the Hybrid Gauss-Least Squares Method with the Venkatakrishnan limiters. The segregated solver employs the SIMPLE algorithm to control the overall solution as follows:

i. Initial all the boundary conditions.
ii. Reconstruct and compute the fields of velocity and pressure gradients.
iii. Solve the momentum equation and get the velocities.
iv. Compute the mass fluxes of all the faces.
v. Solve the pressure correction equation and recalculate the pressure.
vi. Correct the cell velocities and the face mass fluxes.
vii. Update density due to pressure changes.
viii. Go to step iii and continue until convergence or the last iteration flag is reached.
ix. Free all temporary variables and enter the next step.

This study aims to analyze the effects of the gun structure on bubble oscillation and pressure waves in the near field. Therefore, the paper does not consider gravity [5,45,55]. Moreover, at the initial moment, the velocity of the whole domain is zero, and the initial bubble or the chamber of the air gun is filled with high-pressure gas. If not explicitly stated, the coordinate origin of the model is in the center of the air gun or the initial bubble.

## 2.2. Validation

The Zhang equation[4,62] is used here to verify our model, the latest equation for describing the pulsation of the compressible migrating spherical bubble.

$$\left(\frac{C-\dot{R}}{R}+\frac{d}{dt}\right)\left[\frac{R^2}{C}\left(\frac{\dot{R}^2}{2}+\frac{v^2}{4}+H\right)\right]=\ddot{R}R^2+2R\dot{R}^2 \quad (8)$$

Where $R$ denotes the bubble's radius, $H$ indicates the enthalpy, $C$ denotes acoustic velocity, $v$ is the migration velocity, and the dots denote differentiation concerning time.

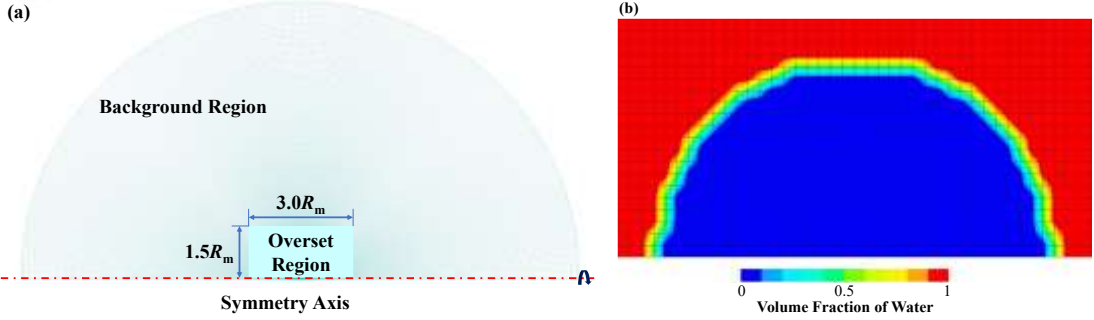

**Fig. 2.** Grid diagram of background and overset region (a), schematic diagram of water-air interface in the overset region (b).

The convergence of the grid has been analyzed in detail, which is tedious, but conclusions are given here. The mesh size of the overset region is better to set less than $R_0/25$, and the value is 0.005 m here, as shown in Fig. 2 (a). At the same time, the area size ought to be larger than $1.5R_m$, where the $R_m$ denotes the maximum bubble radius. For the background region, the mesh size can be 30 times as much as the overset region. Fig. 2 (b). shows the interface between water and gas crosses three grids.

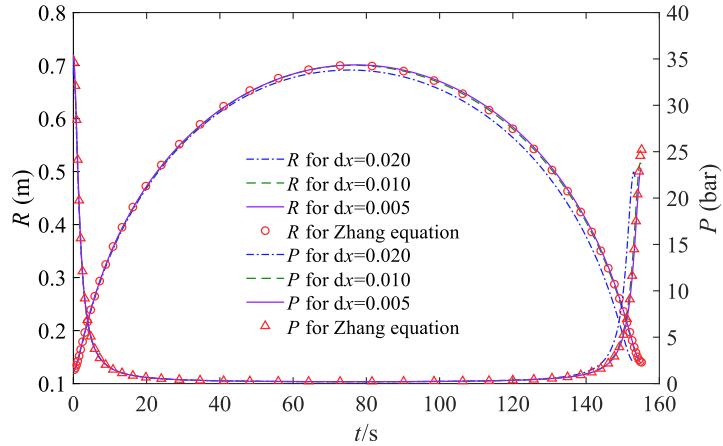

**Fig. 3.** Comparison of the radius $R$ (left axis) and internal pressure $P_{in}$ (right axis) of the bubble from simulations (lines) with Zhang equation (filled symbols). And the d$x$ in the picture is the mesh size of the overset region.

The initial pressure $P_0$ of the bubble is 34.5 bar (500 Psi), and the initial bubble radius $R_0$ is 0.125 m. Fig. 3 compares the Zhang equation radius and internal pressure with the present model in the first bubble period. It agrees well regarding radius and period when the mesh size is 0.005 m. Furthermore, it indicates that the model, including the meshing, is accurate and reliable. Relative pressure contours of the bubble at special moments are presented in Fig. 4, and a solid black line outlines the bubble.

The pictures show that when $t = 10$ ms, the bubble pressure is still more significant than the surrounding water. The bubble keeps expanding rapidly, and the water around the gas is pushed outwards. The pressure in the bubble decreases quickly because of the fast expansion of the bubble volume until the pressure in the bubble is less than the surrounding water, just like $t = 40$ ms. At that moment, the bubble expansion speed begins to reduce. When $t = 80$ ms, the bubble radius and the volume reach their maximum. The bubble interface remains spherical and smooth during both expansion and collision. The pressure of the bubble increases while the bubble volume decreases. At about $t = 154$ ms, the bubble radius has reached the minimum value while the bubble is prepared to rebound and enter the next cycle. All the above results show that the present model is accurate and convincing in solving the bubble pulsation and motion in water, which is considered a two-phase gas-liquid flow.

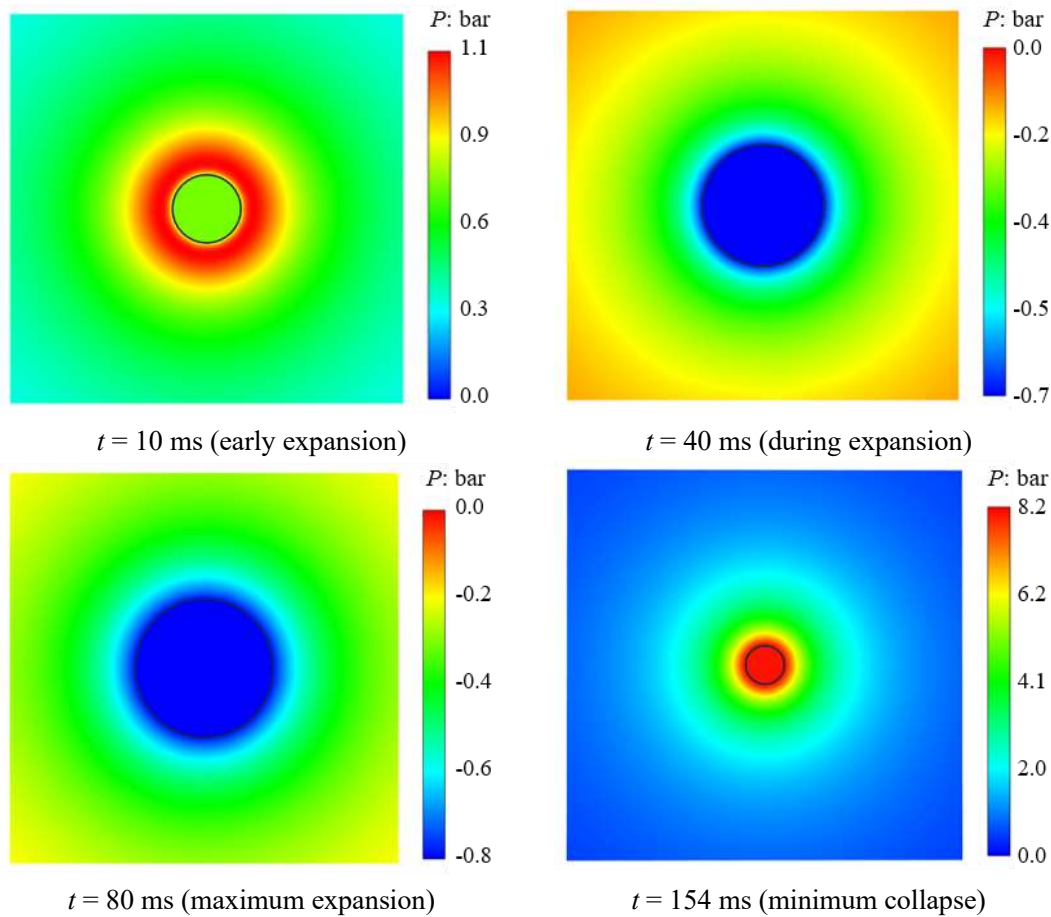

**Fig. 4.** The pressure contours and bubble shapes (solid black line) at certain typical moments (expansion at $t = 10$ ms and 40 ms, maximum expansion at $t = 80$ ms, collapse at $t = 154$ ms).

## 3. Results

### 3.1. Overview

In engineering practices, air guns are commonly designed with one annular opening or four identical openings in the middle[56]. This paper established the air gun model as a cylinder shell with no internal structures, as shown in Fig. 5 (a), which is simplified compared with the practical air gun. $L_{gun}$ and $D_{gun}$ represent the length and diameter of the air gun, $L_{in}$ and $D_{in}$ mean the length and diameter of the air chamber, and $W$ and $H$ indicate the width and height of the opening. The air gun body is assumed to be rigidly fixed, and the interface with water is set as the no-slip wall boundary. The radius of the background region is set to be 1500 m, as in Fig. 5 (b). So, it takes 1 second for the wave to reach the boundary, which is longer than the terminal time.

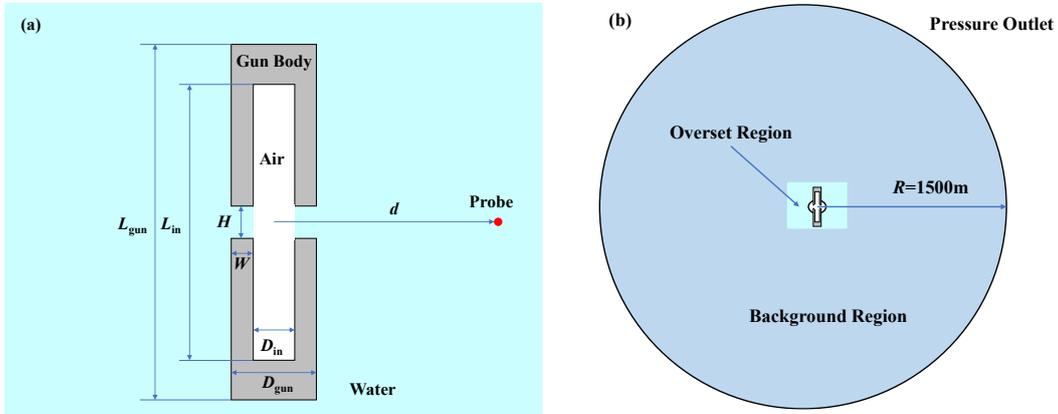

**Fig. 5.** (a) Simplified air gun body with annular opening. (b) The sketch of the computational area: the air gun is nucleated at the center.

We consider a 500 in³ (821.09 cm³) air gun with an initial pressure of 34.5 bar. The parameters of the air gun structure are in Table 1. The gas in the air gun is at rest from the beginning, and the port fully opens when $t = 0$ ms. As the pressure in the firing chamber is much higher than the water, the gas rapidly emerges from the chamber, and the bubble motion can be seen in Fig. 6.

**Table 1** Parameters for the reference group air gun

| | |
|---|---|
| Air gun body length: $L_{gun}$ | 0.190m |
| Air gun body diameter: $D_{gun}$ | 0.304m |
| Chamber length: $L_{in}$ | 0.150m |
| Chamber length: $D_{in}$ | 0.264m |
| Opening height: $H$ | 0.070m |
| Initial pressure: $P$ | 34.5 bar |

The body of the air gun is white in the pictures, and a solid black line outlines the bubble surface. The gas released from the chamber forms a non-spherical bubble at the opening when $t = 2$ ms, and the pressure surrounding the water increases. As the bubble begins expanding, the pressure inside starts to drop. Then, the pressure differential between the chamber and the bubble makes the gas continue to transfer. When $t = 10$ ms, the bubble pressure is below water pressure, and the bubble's expansion is mainly due to the inertia of the water. The bubble radius reaches its maximum and starts to fall when $t = 80$ ms. As the volume of the gas decreases, the bubble pressure increases. When the volume reaches its minimum, it expands again and enters the next period. The abovementioned process involves bubble expansion and mass transfer, making the problem more complicated.

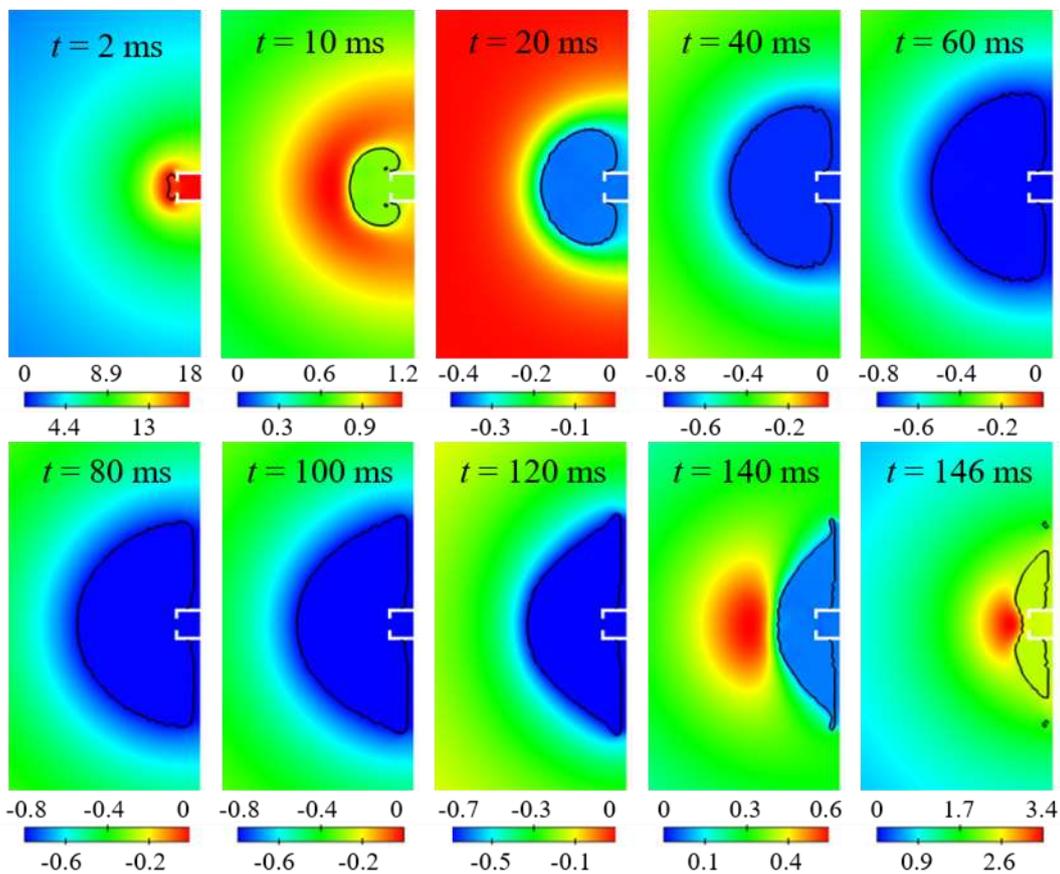

**Fig. 6.** The pressure contours of the air gun bubble at the different typical moments (expansion at $t = 2\sim80$ ms, collapse at $t = 80\sim146$ ms). The color bars depict the relative pressure, and the range varies at different times.

Fig. 7 (a) gives the time series of the volume for the first cycle. The solid black line indicates the volume of the bubble with the same pressure and initial volume in the free field, and the dotted red line shows the volume of the whole gas. Compared with the free-field bubble, the volume of the air gun bubble is slightly smaller because the air gun structure limits the expansion of the bubbles. Meanwhile, not all the gas goes into the bubbles. The pressure waves are related to the volumetric change of the gas for the far field[7,63], and the equation is as follows:

$$\Delta p = \rho_\infty \ddot{V}_b(t - d/C_\infty)/4\pi d \tag{9}$$

Fig. 7 (b) gives the pressure evolution for the probes with different $d$ =1.0 m and 2.0 m. The solid lines are obtained from the present model, and the dashed lines come from the equation (9) while the second derivative of the volume in the equation is calculated based on our model. They are consistent with each other, which shows that the present model is believable and accurate, and the probe $d$ = 1 m belongs to the far field for the model in the case.

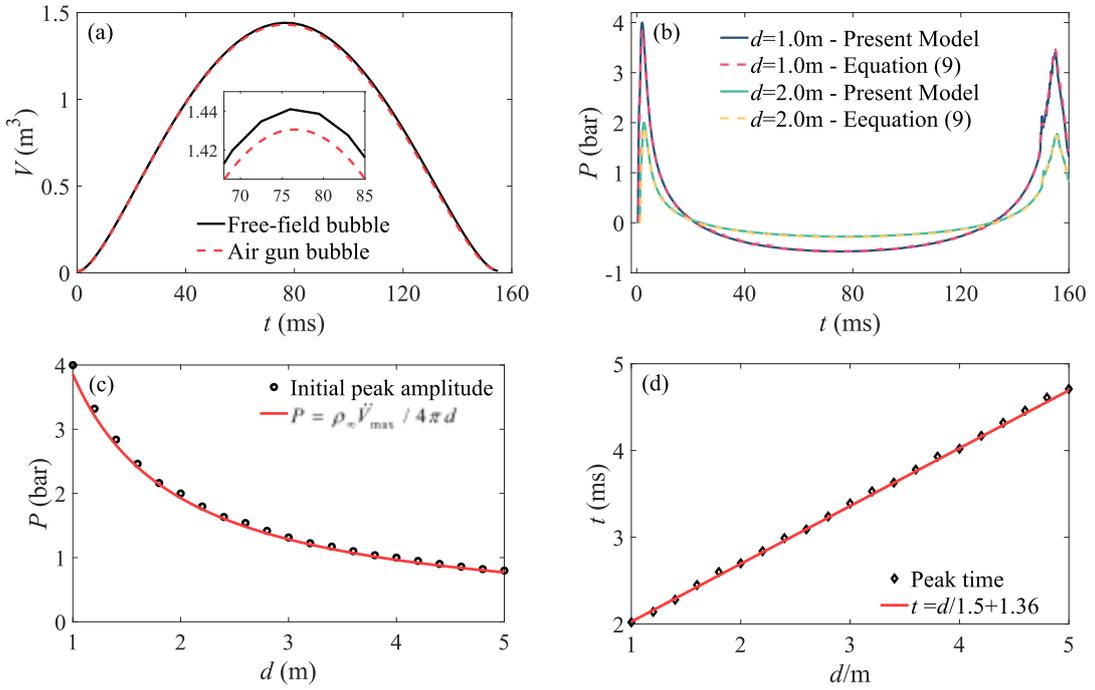

**Fig. 7.** (a) The volume evolution of the free-field bubble and air gun bubble with the same pressure and initial volume, (b) the pressure evolution obtained by the present model and equation (9) for $d$ = 1.0 m and 2.0 m. Initial peak amplitude (c) and peak time (d) for different probes from 1m to 5m, and the red lines are the fitted curve.

Fig. 7 (c) and (d) separately give the amplitude and time of the initial peak for more probes with different $d$. The small black circles in Fig. 7 (c) are the initial peak amplitude for different probes in our model, and they decrease as $d$ increases, which is consistent with the red line. Here, the equation of the red line is the deformed form of the equation (9), and the $\ddot{V}_{max}$ is the maximum value of the second derivative of volume. In Fig. 7 (d), the black diamond indicates the initial peak time for the pressure, and the slope of the red diagonal line is reciprocal of the sound speed (1.5 m/ms), which means that the pressure wave propagates outward at acoustic velocity. To sum up, the pressure waves from air gun bubbles depend upon the second derivative of the gas volume. The initial peak amplitude reduces with distance in the propagation process.

The high frequencies in the pressure waves are usually related to the initial expansion of the gas and the first peak in the initial pressure time series[7]. In Fig. 8, $P^* = \lg P$, a quantity used to better show the initial peak in the pressure space and time series for the absolute pressure $P$. The abscissa represents the axial distance from the original point, and the ordinate represents the time. The pictures have dashed red, dashed-dotted green, and dotted white lines, denoting the interface of the water and air, the side of the opening next to the bubble, and the other side close to the air chamber, respectively.

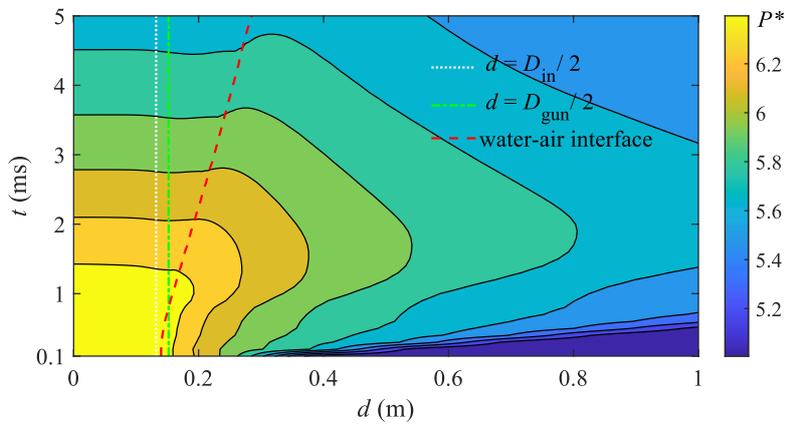

**Fig. 8.** Spatial and temporal distribution of the $P^*$ for the reference group. The parameters of the air gun structure are shown in Table 1.

The formation of the air gun bubble and the propagation of the pressure wave can be seen clearly in the picture. Firstly, the dashed red line represents the speed of the gas expansion, and the intersection with the dashed-dotted green line indicates that the air comes out from the chamber and forms the bubble. After the bubble formation, the initial peak comes and propagates outward. Moreover, on the left of the dashed red line is the air pressure, which is almost uniform from the chamber to the bubble all the time. Such results show that the opening is big enough, and there is no barrier to mass transfer between the air gun bubble and the chamber.

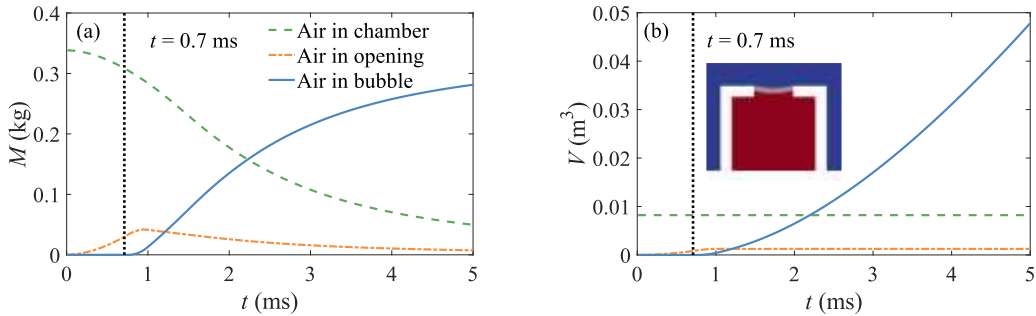

**Fig. 9.** The mass (a) and volume (b) evolution of air in chamber, opening, and bubble in first 5 ms. When $t = 0.7$ ms, the gas comes out from the air gun opening; meanwhile, the bubble starts.

In the previous study[5,25,59], the airflow from the gas chamber into the water is assumed to be isentropic and quasi-static. The gas of the air gun was thought to be only in the chamber and bubble. However, the gas from the chamber goes to the opening first, which is always neglected. In Fig. 9, the dashed, dotted, and solid lines represent the air in the chamber, opening, and bubble. The left figure (a) is the mass, while the right (b) is the volume. As we can see from the pictures, the air in the chamber transfers to the opening first, then comes out from the opening and forms the air gun bubble when $t = 0.7$ ms. Then, the mass transfer increases rapidly and reaches 85% in 5 ms. Such dramatic mass transfer and volume expansion create the first peak of the pressure wave, during which the opening and chamber will significantly impact. Furthermore, their influence will be studied and discussed in detail below.

*3.2. Effect of the Port Opening*

The opening is one of the most critical factors controlling the initial gas release rate, consequently affecting the initial peak. There are two main parameters for the openings: the opening height $H$ and the opening width $W$. Thus, simulations are carried out with different $H$ and $W$. In the first group, the opening height $H$ is set as 0.02 m, 0.03 m, 0.04 m, 0.05 m, 0.06 m, 0.07 m, and 0.08 m. All other parameters of the air gun structure, including the initial pressure, are maintained the same as the reference group in Table 1. The results are plotted in Fig. 10.

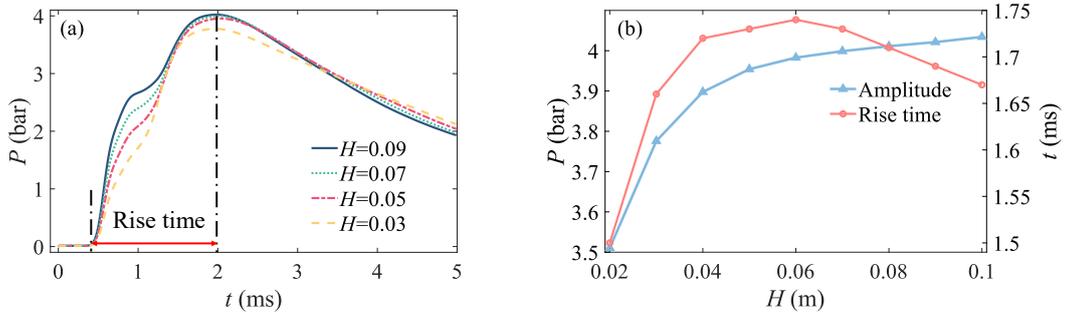

**Fig. 10.** The effects of the port opening height $H$. (a) Pressure wave at the same probe at $d = 1$ m, (b) the initial peak amplitude (left axis) and rise time for different opening height $H$. The other parameters of the air gun body remain the same as the reference group in Table 1.

As we can see from Fig. 10 (a), the pressure changes as the opening height varies from 0.03 m to 0.09 m. During the rise time of the pressure, there is a turning point before which the curve slopes are positively correlated with the size of the opening height $H$. However, the pressure then surges and reaches the peak in a short time almost simultaneously. Fig. 10 (b) gives more details on the amplitude and rise time of the first peak, and the rise time is the duration of the pressure from the start of the rise to the peak. As we can see from the picture, the amplitude of the first peak keeps increasing as the opening height magnifies. In contrast, the rise time increases first and then decreases because the smaller opening height will restrict the gas releasement and reduce the amplitude of the initial peak, and it takes a short time to reach the lower peak. When the opening height is big enough, as $H$ equals 0.06 m, it is not the critical factor to the amplitude, and the rise speed for the peak slows down as the opening increases. However, the gas can be released more quickly, so the rise time drops slowly as the opening increases.

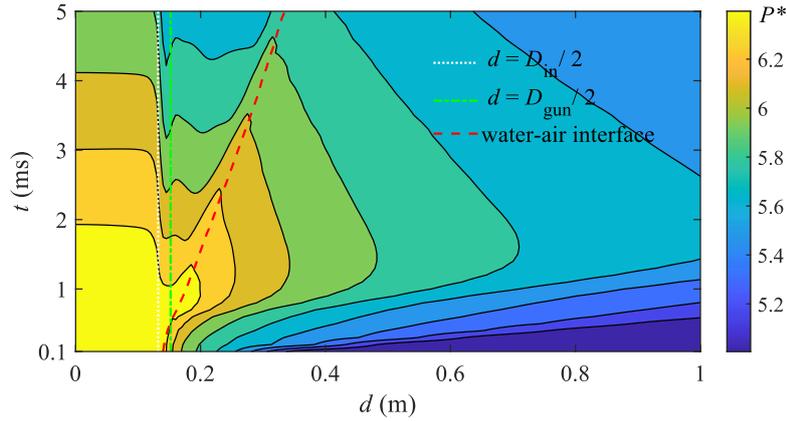

**Fig. 11.** Spatial and temporal distribution of the $P^*$ for $H$ =0.02 m.

The Spatial and temporal distribution of the $P^*$ for $H = 0.02$ m is in Fig. 11. Comparing it with Fig. 8, the pressure in the chamber remains much higher than in the opening and bubble, from which we know that such a small opening height restricts gas flow. Consequently, the initial expansion of the bubble is slower, and the red dotted line is less steep than that of the reference group. In general, the opening height significantly impacts the expansion of the air and pressure wave. However, such an air-release process is always considered isentropic and quasi-static in other literature[7,27,59]. Unlike the quasi-static assumption, the pressure of this model in the chamber and the bubble are not uniform, which is more coincident with the actual situation.

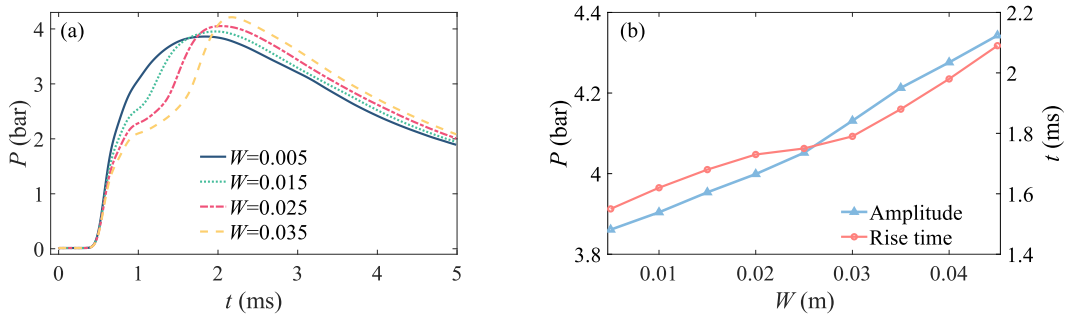

**Fig. 12.** The effects of the port opening width $W$. (a) Pressure wave at the same probe at $d = 1$ m, (b) the initial peak amplitude (left axis) and rise time for different opening width $W$. The other parameters of the air gun body remain the same as the reference group in Table 1.

In the second group, the opening width *W* is set as 0.005 m, 0.001 m, 0.015 m, 0.020 m, 0.025 m, 0.030 m, 0.035 m, 0.040 m, and 0.045 m as it is not wide in actual air gun structural design. The results for port opening widths 0.005m, 0.015 m, 0.025 m, and 0.035 m are shown in Fig. 12 (a), from which we know that with the increase of *W*, the pressure rise gradually slows down, but the peak amplitudes and rise time increase. The curves for different opening widths overlap at the beginning and gradually separate. Moreover, there is no such turning for the pressure wave in the rise if the opening width is small, like *W* = 0.005 m, which indicates that the greater width can also limit the bubble's expansion. The wider *W* expands the air release time and maintains the higher pressure in the chamber for longer. Moreover, the distance from the bubble is shorter when the *W* is greater, resulting in a bigger peak amplitude, which aligns with the facts and is consistent with the previous analysis results.

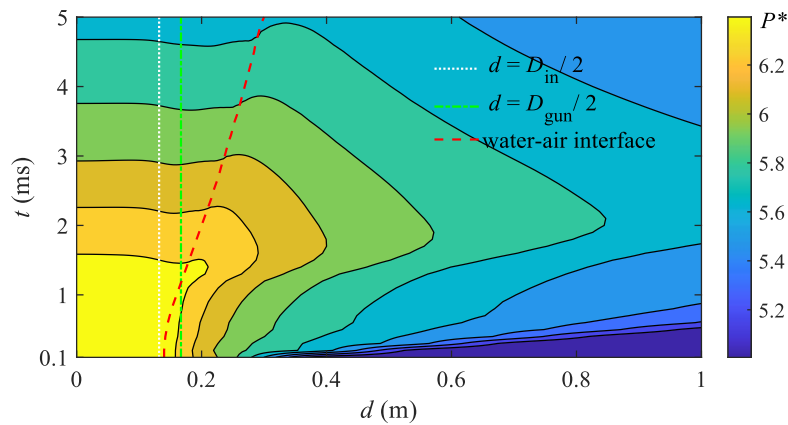

**Fig. 13**. Spatial and temporal and distribution of the pressure for *W* = 0.035 m.

The spatial and temporal distribution of the *P*\* for *W* = 0.035 m is presented in Fig. 13. According to the figure, the air pressure, including the chamber and bubble, is almost uniform, but the wide *W* expands the time of the gas releasement to the bubble. The initial expansion of the gas takes a longer time. As a result, the rise time for the initial peak will also be longer, which is clearly in line with reality. As we can see from Fig. 13, the pressure peak invariably comes after the formation of the bubble, even if the opening is wide. From a particular perspective, in the venting process, if *W* is greater, it will cause a delay in the formation time of bubbles, thereby prolonging the rise time of the peak. This process has always been overlooked in previous research.

From the research above, we conclude that the opening width and height of the air gun will all affect the amplitude and time characteristic of the initial peak. The area of the opening was considered, while the width was ignored by using the mass flow rate equation in previous studies[5,23,25,59], which is not appropriate. Our results indicate that the higher the opening height, $H$, the smaller the width, $W$, and the greater the amplitude. The peak time always comes after the air comes out from the opening as the opening restricts the gas expansion, and the pressure is positively correlated with the gas volume's second derivative. Another important thing: there will be a turning point in the rise time if the opening width is big enough. We assume this is because the gas comes out of the opening, and the gun body no longer restricts the gas release.

*3.3. Effect of the Chamber Shape*

Air guns are always designed to be cylindrical, while the same volume can have different lengths and diameters. In previous articles[7,27,55], there is little research about the effect of chamber shape on the air gun pressure wave, but it plays an essential role as it also decides the difficulty of gas expansion. In this section, specific simulations were carried out to study it, and the length/diameter ratio was introduced to denote the change in the chamber shape. To guarantee the initial volume of the air chamber is 500 in$^3$, the $L_{in}$ and $D_{in}$ must satisfy the equation as follows:

$$\frac{1}{4}\pi D_{in}^2 L_{in} = V_{air} \tag{10}$$

Here are seven groups for comparison, as shown in Table 2. The length of the air gun body $D_{gun}$ equals $D_{in}+2W$, $L_{gun}$ equals $L_{in}+2W$, and $W = 0.005$ m. Furthermore, $S_{open}$ is 0.05 m$^2$ as $H$ changes with $D_{in}$ to keep $S_{open}$ unchanged. The other parameters are the same as the reference group.

**Table 2** Parameters of the groups for different $L_{in}/D_{in}$ (unit: m)

| $L_{in}/D_{in}$ | 0.5 | 1 | 1.5 | 2 | 2.5 | 3 | 3.5 |
|---|---|---|---|---|---|---|---|
| $L_{in}$ | 0.137 | 0.217 | 0.284 | 0.344 | 0.399 | 0.451 | 0.500 |
| $D_{in}$ | 0.273 | 0.217 | 0.189 | 0.172 | 0.160 | 0.150 | 0.143 |
| $H$ | 0.058 | 0.073 | 0.084 | 0.092 | 0.099 | 0.105 | 0.111 |

The initial peak pressure curves are plotted in Fig. 14 (a) for the probe, which is 1 m far from the center point of the air gun. Fig. 14 (b) gives the initial peak amplitude (left axis) and rise time (right axis). The results above show that the smaller the length/diameter ratio, the faster and higher the initial pressure wave rises. Furthermore, their initial slope is the same as they have the same opening area. As the gas is released, the expansion of the gas is gradually confined by the gas chamber. The greater the length/diameter ratio, the slower the bubble expands, remarkably affecting the initial peak and rise time.

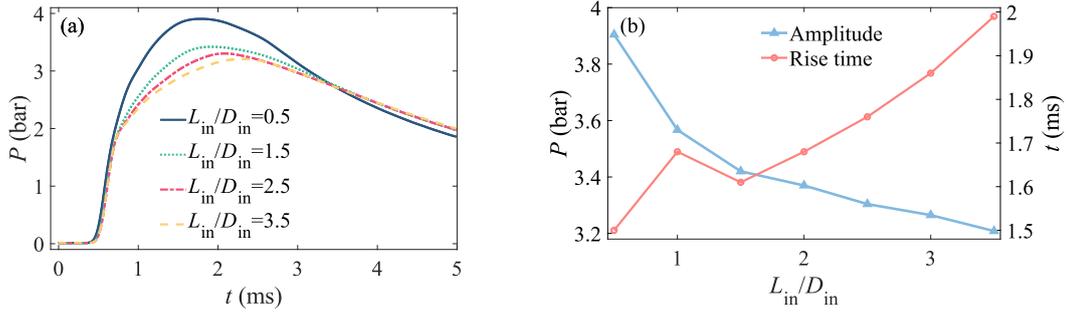

**Fig. 14.** The effects of the length/diameter ratio. (a) Pressure wave at the same probe at $d = 1$ m, (b) the initial peak amplitude (left axis) and rise time for different $L_{in}/D_{in}$. The other parameters of the air gun body, except the opening height, are the same as the reference group in Table 1.

The spatial and temporal distribution of the $P^*$ for $L_{in}/D_{in} = 3.5$ is shown in Fig. 15. According to the figure, the air pressure, including the chamber and bubble, is almost uniform. However, the larger value of the length/diameter ratio expands the air release time to bubble, which means the initial expansion of the air takes longer. Consequently, the rise time for the initial peak will also be longer. It also proves that the throttling effect is less apparent as its opening area is 0.05 m$^2$. However, the elongated chamber significantly affects the gas release rate, and it could be explained by considering the spatial distribution of the pressure in the air gun firing chamber, which shares the same view as Watson et al. (2019).

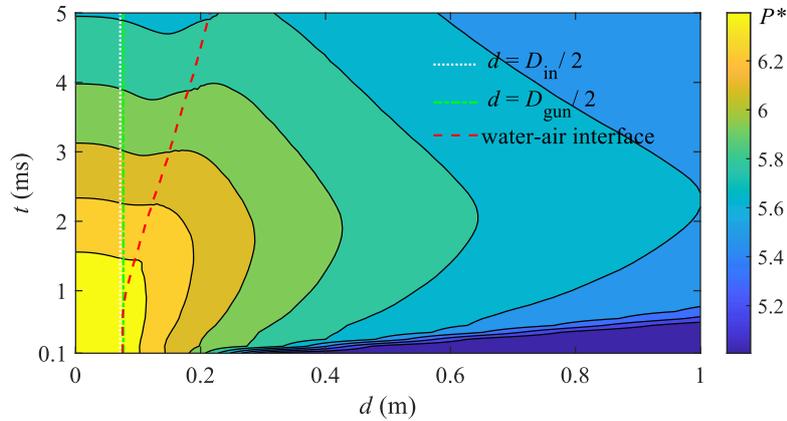

**Fig. 15.** Spatial and temporal distribution of the pressure $P^*$ for $L_{in}/D_{in} = 3.5$.

Through the above analysis, we know that the pressure wave from the air gun mainly depends on the acceleration of the volume expansion of the gas. Besides the air gun depth and pressure[7,23,59], the air gun body affects the venting process, especially the initial stage, which is crucial to the initial peak. The opening height, width, and chamber shape all significantly impact the expansion of the air and pressure wave, and such difference makes the characteristics of the air gun source.

**4. Conclusion**

This paper establishes the compressible bubble model, and the comparisons with the Zhang equation show that the model is accurate. The air gun bubble model with a complicated body is established based on the multiphase flow model, and the process of gas passing through the air gun opening is considered. The results show that the pressure wave from the air gun is mainly proportional to the expansion acceleration of the whole gas. The amplitude of the first peak in the field depends on the distance if the parameters of the air gun are unchanged. Moreover, the amplitude of the first peak reduces as distance increases and is the inverse of the distance while the rise time remains the same. Therefore, the slope of the first peak depends upon the initial pressure and firing depth in the water for certain ready-made air guns.

The acceleration of volume expansion of the whole gas is the critical factor that controls the pressure wave ejected from the air gun. Our results show that the body of the air gun, such as the chamber shape and opening, affects the expansion of the whole gas. The larger the opening is, and the faster the gas is released first, the shorter the rise time for the initial peak is. The larger the chamber length/diameter ratio, the slower the gas is released and the longer the rise time for the initial peak is. However, the smaller opening and larger chamber length/diameter ratio will also lead to mass throttling, reducing the peak amplitude. The results above have great significance for the newly designed air gun.

**Acknowledgments**

This work is supported by the Finance Science and Technology Project of Hainan Province (No. ZDKJ2021020), the National Key R&D Program of China (No. 2022YFC2803500), the National Natural Science Foundation of China (Grant No. 52371313 and 52088102), and Hainan Special Ph.D. Scientific Research Foundation of Sanya Yazhou Bay Science and Technology City (HSPHDSRF-2022-06-001).